\documentclass[12pt]{article}
\pdfoutput=1
\usepackage{graphicx}
\usepackage{bbm}
\usepackage{bbold}
\usepackage{authblk}
\usepackage{amsfonts}
\usepackage{dsfont}
\usepackage{cite}

\begin{document}

\title{A quantum protective mechanism in photosynthesis}

\author[1]{Adriana Marais\thanks{Corresponding author: adrianamarais@gmail.com}}
\author[1]{Ilya Sinayskiy}
\author[1]{Francesco Petruccione}
\author[2]{Rienk van Grondelle}
\affil[1]{Quantum Research Group, School of Chemistry and Physics,
University of KwaZulu-Natal, Durban, 4001, South Africa and National Institute for Theoretical Physics, KwaZulu-Natal, South Africa}
\affil[2]{Institute for Lasers, Life and Biophotonics, Faculty of Sciences, VU University Amsterdam, De Boelelaan 1081, 1081 HV, Amsterdam, The Netherlands}

\date{Dated: \today}

\maketitle

\newpage
\begin{abstract}
Since the emergence of oxygenic photosynthesis, living systems have developed protective mechanisms against reactive oxygen species. During charge separation in photosynthetic reaction centres, triplet states can react with molecular oxygen generating destructive singlet oxygen. The triplet product yield in bacteria is observed to be reduced by weak magnetic fields. Reaction centres from plants' photosystem II share many features with bacterial reaction centres, including a high-spin iron whose function has remained obscure. To explain observations that the magnetic field effect is reduced by the iron, we propose that its fast-relaxing spin plays a protective role in photosynthesis by generating an effective magnetic field. We consider a simple model of the system, derive an analytical expression for the effective magnetic field and analyse the resulting triplet yield reduction. The protective mechanism is robust for realistic parameter ranges, constituting a clear example of a quantum effect playing a macroscopic role vital for life.

\end{abstract}
Over the past few billion years, a class of living systems has perfected a method of synthesising organic compounds from carbon dioxide and an electron source using the sunlight energy continuously incident on the surface of the Earth: the process is known as photosynthesis. The oxygenation of the early atmosphere through oxygenic photosynthesis using water, facilitated efficient cellular respiration with O$_2$ as the electron acceptor and the subsequent development of multicellular life \cite{raymond06}. Furthermore, the conversion of O$_2$ into ozone by ultra-violet light in the upper atmosphere provided a protective layer beneath which life has flourished \cite{falkowski06}. The benefits of oxygenation, however, came at a price: reactive oxygen species are toxic to living cells \cite{halliwell99}, and the evolution of protective mechanisms against damaging oxygen species became essential for survival.\\
\\
Anoxygenic purple bacterial reaction centers (RCs) were the first to be biochemically isolated and characterised \cite{deisen85}, and are simpler in structure than those of evolutionarily more recent higher plants. Electron transfer is initiated in purple bacterial RCs when a photoinduced electronic excitation is transferred through a pigment-protein antennae system to a pair of bacteriochlorophyll molecules, P \cite{baha04}. A single electron is then transferred sequentially down the active $A$-branch \cite{novod04}, finally reducing the acceptor Q$_B$ in the $B$-branch. The resulting stable charge-separated state provides energy for subsequent chemical conversions  \cite{blank95}.\\
\\ 
If, however, electron transfer to the ubiquinone molecule Q$_A$ in the active branch is blocked, either through natural over-reduction of the pool of quinones or in the laboratory by chemically reducing Q$_A$, the lifetime of the spin-correlated radical pair P$^+$H$_A^-$ increases to 10-20 ns \cite{ogrod82,schenck82,chidsey84}. During this time the interaction of each radical with its respective hyperfine environment results in oscillations between the initial singlet state and the triplet state of the pair. Subsequent reactions depend on the instantaneous spin multiplicity of the pair, and charge recombination results in the formation of either singlet or triplet products.\\
\\
In the RC of the purple bacterium \textit{Rhodobacter sphaeroides} with reduced acceptor Q$_A$, the yield of triplet products is observed to be lowered by weak external magnetic fields \cite{hoff77,blank77}. This effect can be accounted for by the theory of chemically induced magnetic polarisation \cite{kaptein72}, whereby a sufficiently strong external magnetic field decreases the population of triplet states, due to the dependence of their energies on the magnitude of the field, which results in reduced singlet-triplet conversion.\\
\\
The formation of high energy triplet states involving (bacterio)chlorophyll are of potential danger in all photosystems, since they can react with molecular oxygen generating highly reactive singlet oxygen \cite{schweitzer03}, which is damaging to photooxidation reactions, bleaching pigments and bringing about protein inactivation and lipid peroxidation \cite{nooden03} and can be damaging to biological material in general \cite{krinsky79}; being implicated in aging and disease \cite{ames93,shigenaga94,stadtman97}. A more recent experiment has demonstrated that the yield of singlet oxygen in carotenoidless bacterial photosynthetic RCs, and as a consequence the stability of the RC protein, are strongly magnetic field-dependent \cite{liu04}.\\ 
\\
The presence of a carotenoid in the RCs of anoxygenic purple bacteria \cite{thornber80}, with the principal function of quenching the energy of triplet products \cite{dewinter99}, suggests that singlet oxygen formation is worth protecting against, even in low-oxygen environments. In the oxygen-rich environment of the water-splitting photosystem II RC, protective mechanisms are even more crucial  \cite{newell99}. Evidence suggests that radical pair recombination contributes significantly to singlet oxygen production in the chloroplast, which in turn inhibits the repair of light-induced damage in photosystem II \cite{hakala11}.\\
\\
Photosystem II RCs in higher plants share many structural and functional features with the bacterial RC, reflecting evolution from a common ancestral RC \cite{allen11,rutherford12}. Amongst these similarities is an Fe$^{2+}$ ion, a spin-2 iron ion, positioned between the two ubiquinone molecules in both types of RCs \cite{blank95}, see Fig. 1. The ion has been postulated to play a structural and/or energetic role in electron transfer \cite{feher99}. However, native dynamics can be restored in Fe$^{2+}$-depleted RCs under certain conditions, including when the paramagnetic ion is replaced by diamagnetic Zn$^{2+}$ \cite{liu91}.\\
\\
This work is based on a series of experimental observations that the Fe$^{2+}$ ion has an effect on radical pair reactions in blocked bacterial reaction centres. Early experiments showed an increase in the relative magnetic field effect in Fe$^{2+}$-depleted RCs \cite{hoff77,blank77}, and more recently in an experiment by Kirmaier et al. \cite{kirm86}, an increased yield of triplet products was observed in Fe$^{2+}$-depleted RCs relative to native RCs.\\
\begin{figure}
\centering
  \includegraphics[scale=0.4]{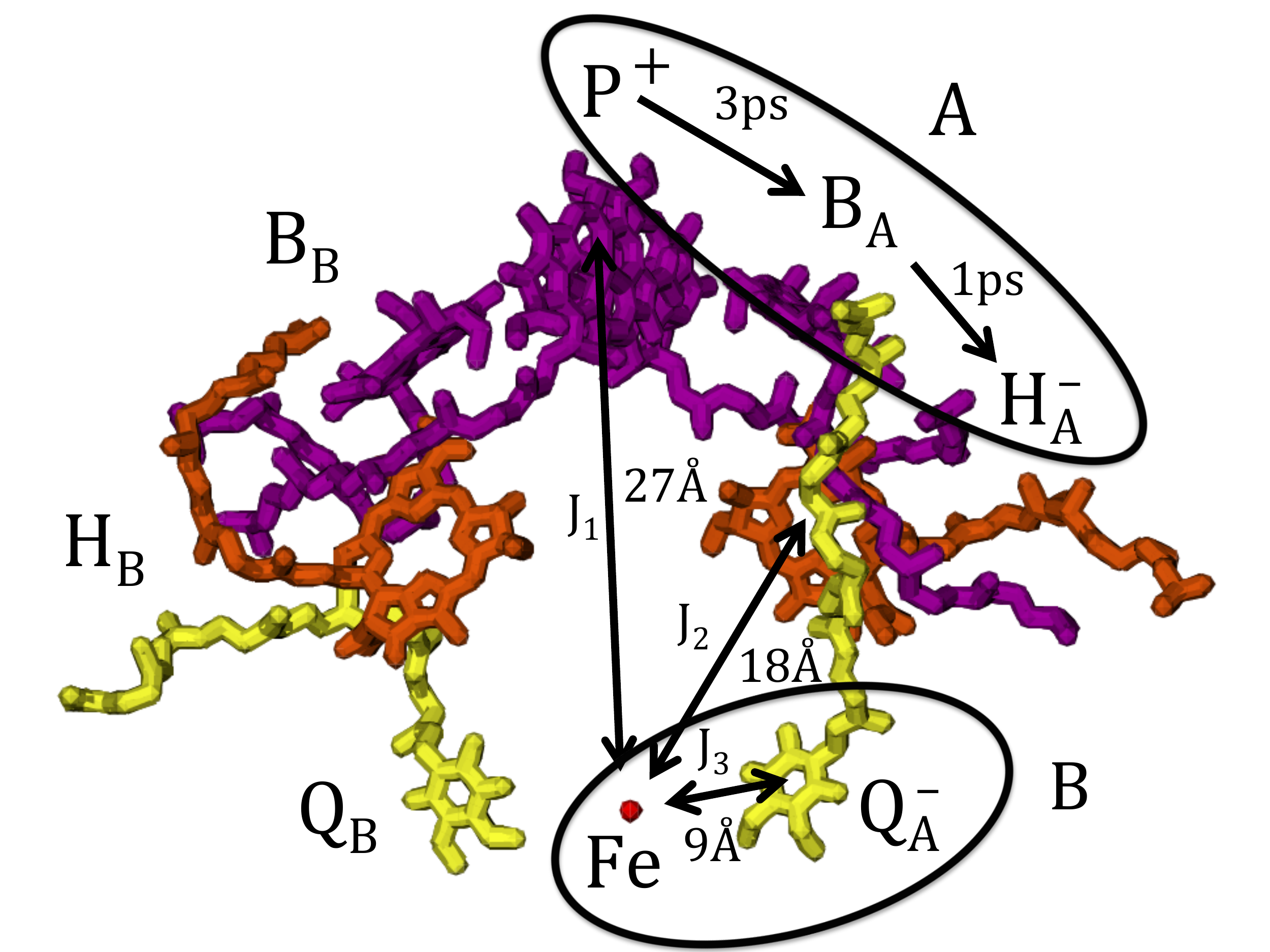}
    \caption{The three-dimensional structure of the cofactors of the reaction center from \textit{Rhodobacter sphaeroides} R-26 carotenoidless strain. Shown are the bacteriochlorophyll \textit{a} dimer P and the bacteriochlorophyll \textit{a} monomers B$_A$  and B$_B$ (purple), the bacteriopheophytin \textit{a} molecules H$_A$  and H$_B$ (orange), the ubiquinone molecules Q$_A$  and Q$_B$ (yellow), and the iron Fe (red). In RCs at room temperature with the ubiquinone removed or reduced, the lifetime of the the radical pair P$^+$H$_A^-$ formed on the active $A$-branch is 10-20 ns and singlet-triplet mixing is observed.}
\end{figure}
\\ 
Based on these observations, we propose a mechanism via which such an effect could occur. We propose that the high-spin Fe$^{2+}$ ion generates an effective magnetic field, thereby serving the protective function of reducing the triplet product yield in the purple bacterial RC under conditions when forward electron transfer is blocked.  Evidence of quantum coherence in primary energy transfer in photosynthesis \cite{engel07,panit10,collini10} has resulted in the successful application of quantum models of environment-assisted energy transfer to the biological process \cite{mohseni08,caruso09,semiao10,asadian10,sarovar10,mulken10,scholak11,sinay12,marais13}. Here the direct role of spin in the proposed protective mechanism during the subsequent process of primary charge separation constitutes a clear example of a quantum effect of macroscopic importance to a living organism. 
\section*{Results}
\subsection*{The effective magnetic field generated by a fast-thermalising spin}
A spin-correlated radical pair consists of two atoms, molecules or ions each with an unpaired valence electron or open electron shell, where the spins of the electron, or electron hole, add to give a total spin angular momentum of 1 (triplet state), or 0 (singlet state). The three degenerate triplet states, $|t_0\rangle$ and $|t_{\pm}\rangle$, are energetically separated from the singlet state, $|s\rangle$, by the magnitude of the interaction between the two radicals, which depends on their separation. The interaction of each of the radicals with a different hyperfine environment induces singlet-triplet mixing. A magnetic field modifies this mixing, and when its magnitude exceeds that of the hyperfine interaction, the states $|t_{\pm}\rangle$ are separated from the remaining states' energy levels due to the Zeeman interaction. The spin-dependent reaction kinetics thus become magnetic field dependent \cite{steiner89}.\\
\\
According to Kramers' theorem \cite{kramers30}, energy level degeneracies of integer spin systems are lifted by electrostatic fields resulting from local charge distributions \cite{ball71}, while those of non-integer spin systems remain at least doubly degenerate. For systems in the former category, including the Fe$^{2+}$ ion, direct spin-lattice coupling and resulting energy level transitions result in the rapid thermalisation of the spin system \cite{shiren62}, relative to the Kramers' doublets in the latter, which include the pair of radicals constituting P$^+$H$_A^-$.\\
\\
The procedure of adiabatic elimination \cite{breuer02} can be applied to quantum systems with dynamics occurring on widely separated timescales. The effective dynamics for a `slow' system of interest are derived by assuming that a `fast' system process is completed on a timescale where the `slow' system is static. We will now show that the reduced dynamics for a radical pair interacting with a fast-thermalising spin are analogous to the dynamics of a radical pair in an external magnetic field.\\
\\ 
We divide the system into two subsystems: subsystem $A$ consisting of the radical pair ($A_1$ and $A_2$), and subsystem $B$ consisting of the high-spin ($B_1$) and strongly coupled spin-1/2 ($B_2$). The Hamiltonian for subsystem $B$ is given by 
\begin{equation}
H^{B} = \textbf{S}^{B_1}\cdot\textbf{J}_3\cdot\textbf{S}^{B_2} + D\left((S^{B_1}_z)^2-S(S+1)\mathds{1}/3)\right) + E\left((S^{B_1}_x)^2 - (S^{B_1}_y)^2\right),
\end{equation}
while the Hamiltonian $H^{AB}$ for the total system is given by 
\begin{equation}
H^{AB} = \textbf{S}^{A_1}\cdot\textbf{J}_1\cdot\textbf{S}^{B_1} + \textbf{S}^{A_2}\cdot\textbf{J}_2\cdot\textbf{S}^{B_1} + H^A + H^B.
\end{equation}
Here, $\textbf{S}$ are spin operators with standard commutation relations for arbitrary values $S$ of spin and $\textbf{J}_i$ are the interaction tensors for the pairs of spins. The zero field splitting parameters $D$ and $E$ quantify the lifting of the spin state degeneracy as a result of indirect effects of the local electrostatic field, which can cause a non-spherical electron distribution within systems with spin $S>$1/2.\\
\\
The use of a spin Hamiltonian is justified if the $2S+1$ spin ground states are far removed in energy from other sets of states with higher orbital angular momentum. This is the case in the bacterial RC, since the first excited quintet of spin states are determined to lie 490 K above the ground state quintet \cite{feher99}.\\
\\
The dynamics of the total system are given by the following master equation
\begin{equation}
\dot{\rho}^{AB} = (\mathcal{L}^{A} + \mathcal{L}^{B})\rho^{AB},
\end{equation}
where the superoperators $\mathcal{L}$ are the standard dissipation superoperators from Dicke-like models \cite{dicke54}, defined by their action on density matrix $\rho$ as
\begin{eqnarray}
\mathcal{L}^{A}\rho&=&-i[H^{AB},\rho],\\
\mathcal{L}^{B_1}\rho&=&\gamma(\langle n\rangle+1)(S^{B_1}_{-}\rho S^{B_1}_{+}-\frac{1}{2}\{S^{B_1}_{+}S^{B_1}_{-},\rho\})\nonumber\\
&& +\gamma\langle n\rangle(S^{B_1}_{+}\rho S^{B_1}_{-}-\frac{1}{2}\{S^{B_1}_{-}S^{B_1}_{+},\rho\}),
\end{eqnarray}
where $S_{\pm}^{B_1}$ are the lowering and raising spin operators for spin-$S$ subsystem $B_1$, and $\gamma,\langle n\rangle$ are the spontaneous emission coefficient and average number of thermal photons at the frequency of transition, respectively.\\
\\
Assuming that the high-spin subsystem $B_1$ thermalises fast and is strongly coupled to the spin-1/2 subsystem $B_2$ compared to other system processes and couplings, i. e. $\gamma,\gamma\langle n\rangle, |J_3| \gg |J_1|,|J_2|$, which is a good approximation for the bacterial RC \cite{calvo02}, we apply the method of adiabatic elimination of subsystem $B$, in this approximation only the single spin-$S_{B_1}$ being dissipated. Using the projection operator technique \cite{breuer02}, projection operator $P$ on the relevant part of the system is defined as
\begin{equation}
P\rho^{AB} = \mathrm{Tr}_B(\rho^{AB})\otimes \rho_{ss}^B,\\
\end{equation}
where $\rho^B_{ss}$ is the steady state solution to the equation $\dot{\rho}^{B}_{ss}=\mathcal{L}^{B}\rho^{B}$, i. e. $\mathcal{L}^{B}\rho^{B}_{ss} = 0$. The dynamics for reduced system $A$ where $\rho^A = $Tr$_B(\rho^{AB})$ are then given by a van Kampen-like expansion \cite{breuer02}
\begin{equation}
\dot{\rho}^{A} = \langle\mathcal{L}^{A}\rangle_B\rho^{A} + \int\limits_0^\infty\mathrm{d}\tau(\langle\mathcal{L}^{A}\mathrm{e}^{\mathcal{L}^B\tau}\mathcal{L}^{A}\rangle_B - \langle\mathcal{L}^{A}\mathrm{e}^{\mathcal{L}^B\tau}\rangle_B\langle\mathcal{L}^{A}\rangle_B)\rho^A, 
\end{equation}
where the operators on subsystem $B$ have been replaced with their expectation values with respect to the steady state $\rho^{B}_{SS}$ reached by the fast-relaxing subsystem $B$
\begin{equation}
\langle O\rangle_B = \mathrm{Tr}_{B}(O\rho^{B}_{SS})
\end{equation}
for operator $O^B$ on subsystem $B$.\\
\\
Taking into account that $\langle\vec{S}\rangle_B\neq 0$, and that the effective radical pair mechanism is therefore a first order effect, and we consider the dynamics for the reduced system $A$ to first order only
\begin{equation}
\dot{\rho}^{A} = \langle\mathcal{L}^{A}\rangle_B\rho^{A}.
\end{equation}
The expectation values for the spin operators $S_x$ and $S_y$ are zero with respect to the steady state $\rho^B_{ss}$, satisfying $\dot{\rho}^{B}_{ss}=\mathcal{L}^{B}\rho^{B}_{ss}=0$ for spin-$S$ subsystem $B$, and are also zero in the case where $J_3=0$. The reduced dynamics of subsystem $A$ are therefore given by
\begin{eqnarray}
\dot{\rho}^{A}&=&-i[\langle H^{AB}\rangle,\rho^A]\\
&=&-i[\langle S_z^B\rangle(J_{1z} S_z^{A_1} + J_{2z} S_z^{A_2}) + H^A,\rho^A].
\end{eqnarray}
It be seen from the form of Eq. (11) that the reduced dynamics of the radical pair, in the limit where subsystem $B$ is adiabatically eliminated, are mathematically equivalent to that of system $A$, the radical pair, in a magnetic field in the $z$-direction, with the effective magnetic field strength for each of the radicals $A_1$ and $A_2$ given by
\begin{equation}
B_{\mathrm{eff}}^{j} = \frac{J_{j}\langle S_z^B\rangle}{\mu_{B}g_{j}}
\end{equation}
where $\mu_B$ is the Bohr magneton and $g_{j}$  are the g-values for each of the radicals $A_j$. In the context of the NMR, the analogous effective magnetic field generated by a fast-relaxing spin gives rise to the so-called Knight shift \cite{carr67}. Note that since the expectation value $\langle S_z^B\rangle<0$, the magnetic field effect on a radical pair in the presence of an external magnetic field is reduced by the effective magnetic field generated by the fast-thermalising spin for positive $J_j$.\\
\begin{figure}
\centering
  \includegraphics[scale=0.5]{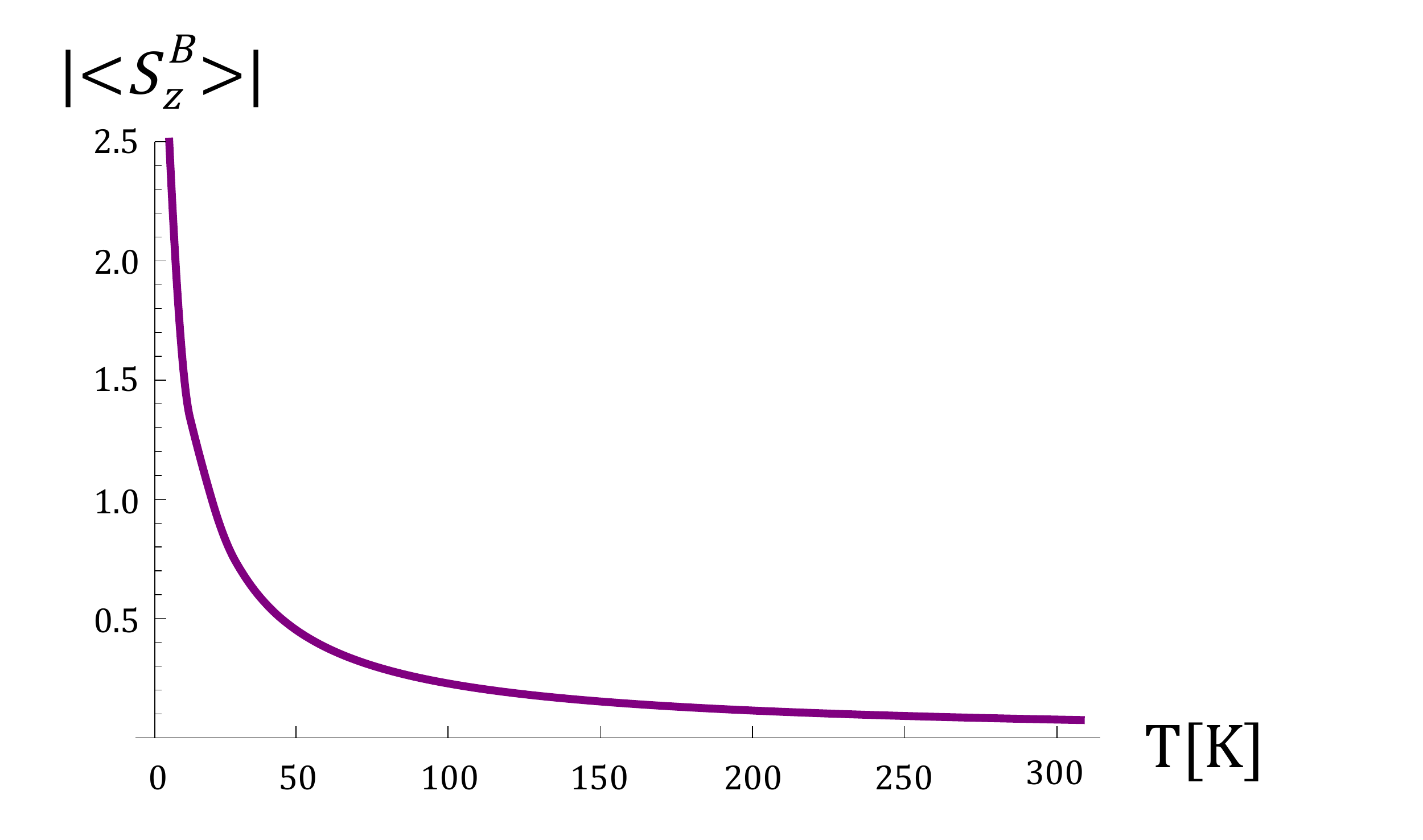}
    \caption{The dependence of the magnitude of the expectation value of $S^{B}_z$ for the Fe$^{2+}$ ion and strongly coupled Q$^-$ on the effective temperature corresponding to the average number of thermal photons at the frequency of transition $\langle n\rangle$.}
\end{figure}
\\
As the spontaneous emission coefficient $\gamma$, which parametrises magnetic dipole transitions induced by thermal microwave photons for the Fe$^{2+}$ ion, is increased for a fixed temperature, a maximum value off $\langle S_z^{B}\rangle$ is obtained. Setting the value of the coefficient to $\gamma = 1.5\times10^5$ G, results in obtaining the maximum value of $\langle S_z^{B}\rangle$ to two significant figures for all considered temperatures. In Fig. 2, the dependence of the expectation value of $|\langle S_z^{B}\rangle|$ on the effective spin temperature of the strongly coupled Fe$^{2+}$ and Q$^-_A$ spin system is plotted, where the average number of thermal photons $\langle n\rangle$ as a function of temperature is given by
\begin{equation}
\langle n\rangle = \frac{1}{\mathrm{e}^{\hbar\omega/k_BT}-1}.
\end{equation}
(In this paper, we use Gauss as the energy ($E$) units, i. e. $E/g\mu_B$ where $\mu_B$ is the Bohr magneton and for g=2 in units with $\hbar=1$, 1 G = 0.0028 ns$^{-1}$.)
\subsection*{Triplet yield reduction}
We now investigate the extent to which the effective magnetic field generated by the Fe$^{2+}$ and Q$^-_A$ spin system reduces the triplet product yield when forward electron transfer is blocked in the bacterial RC.\\
\\ 
We consider the following simple model with spin Hamiltonian $H_N$ given by
\begin{equation}
H_N = J\textbf{S}_1\cdot\textbf{S}_2 +  \sum\limits_{i}^{N_1}A_i\textbf{I}_i\cdot\textbf{S}_1 + \sum\limits_{j}^{N_2}A_j\textbf{I}_j\cdot\textbf{S}_2 + \omega(S_{z_1}+S_{z_2}), 
\end{equation}
where $J$ is the interradical exchange interaction, $A_j$ are the isotropic hyperfine couplings with spin-1/2 nuclei on each of the radicals P$_A^+$ and H$_A^-$, $\omega$ is the electron Larmor frequency in the effective magnetic field generated by the Fe$^{2+}$ and Q$^-_A$ spin system, and $N_1+N_2=N$ is the total number of coupled nuclei.\\
\\
In a quantum master equation approach to radical pair reactions, which conserves probability unlike the Liouville equations typically employed in this setting, we use the formalism introduced in ref. \cite{kominis09}, with so-called `shelving states'  \cite{gauger11} $|S\rangle$ and $|T\rangle$ to represent the singlet and triplet products that are spin-selected from the initial electronic singlet $|s\rangle$ or triplet state $|t_{0,\pm}\rangle$ of the radical pair. Mathematically, we use the direct sum to extend the Hilbert space of the radical pair to include the shelving states $|S\rangle$ and $|T\rangle$ as extra basis vectors. We define the four Lindblad jump operators implementing these decay events as 
\begin{eqnarray}
P_{s}&=&|S\rangle\langle s|\otimes\mathds{1}_B,\\
P_{t_{0,\pm}}&=&|T\rangle\langle t_{0,\pm}|\otimes\mathds{1}_B.
\end{eqnarray}
The state of the bath consisting of $N$ nuclei which is coupled to radicals 1 and 2, $\rho_B$, is given by
\begin{equation}
\rho_B = \prod\limits_{i=1}^{2}\frac{1}{Z_i}\mathrm{e}^{-\beta\alpha_iS_{z_i}},
\end{equation}
where $Z_i$ is the partition function of the corresponding bath, 
\begin{equation}
Z_i = \sum\limits_{j_i=0}^{N_i/2}\sum\limits_{m_i=-j_i}^{j_i}\nu(N_i,j_i)\langle j_i,m_i|\mathrm{e}^{-\beta\alpha_iS_{z_i}}|j_i,m_i\rangle,
\end{equation}
$\beta$ is the inverse temperature, $\nu(N_i,j_i)$ denotes the degeneracy of the environmental spins, $|j_i,m_i\rangle$ denote the well-known eigenvectors of the angular momentum operator $S_i$, and $S_{z_i}$ are the collective operators \cite{sinay12,marais13,dicke54,breuer02}.\\
\\
The master equation is given in terms of the rates of recombination to singlet and triplet products, $k_S$ and $k_T$, by
\begin{eqnarray}
\dot{\rho}&=&-\frac{i}{\hbar}[H_N,\rho]\ +\ k_S\left(P_s\rho P_s^{\dagger} - \frac{1}{2}(P_s^{\dagger}P_s\rho+\rho P_s^{\dagger}P_s)\right)\nonumber\\
&& +\ k_T\left(\sum\limits_{t=t_0,t_{\pm}}P_t\rho P_t^{\dagger} - \frac{1}{2}(P_t^{\dagger}P_t\rho+\rho P_t^{\dagger}P_t)\right),
\end{eqnarray}
where $H_N$ is given in Eq. (14). The singlet and triplet product yields, $\phi_S(t)$ and $\phi_T(t)$, are given by the populations of the shelf states as functions of time.\\
\\
Based on values determined in the literature, mostly a few decades ago, we define a broad yet reasonable range of parameters over which the extent and importantly the robustness of the protective effect is analysed:\\
\\
While singlet and triplet yields are often evaluated at infinite times in theoretical models in the literature, the lifetime of the radical pair is in fact 10-20 ns in blocked RCs \cite{ogrod82,schenck82,chidsey84}, and we therefore consider times in this interval.\\
\\
Electron-nuclear hyperfine interactions lie in the approximate range between 1 and 100 G \cite{timmel98}.  While theoretical modelling and ENDOR measurements of the relevant pigment radicals in solution determine values in the range between 1 and 20 G (see refs \cite{ogrod82,werner78} and refs therein), these values may be higher in the rigid protein environments of biological RCs, and therefore, we consider hyperfine coupling strengths between each radical and its respective hyperfine environment to be in the range between 1 and 100 G.\\
\\
Each radical interacting with the strongly coupled Fe$^{2+}$ and Q$^-$ spin system experiences an effective magnetic field $B_{\mathrm{eff}}^{j}$. This effective field depends on the coupling $J_{1,2}$ with the spin system which in turn depends on the distance of each radical from the ion (see Fig. 1), and also on the effective spin temperature of the strongly coupled spin system (see Fig. 2).\\
\\
The Fe$^{2+}$ ion is positioned at a distance of 18.3 $\textrm{\AA}$ from the H$^-$ radical, while the distance between the Fe$^{2+}$ ion and the P$^+$ radical is 28.8 $\textrm{\AA}$, a distance at which the Heisenberg exchange interaction becomes negligible. However, evidence of coupling of a dipole-dipole nature between the primary electron donor and the Fe$^{2+}$ ion in both bacterial and photosystem-II RCs has been reported \cite{hirsh93}. Estimates for such couplings in RCs range from around 1 to 50 G \cite{calvo02,norris82,ogrod82,werner78}.\\
\\
Pach{\'o}n and Brumer \cite{pachon11} argue that long-lived coherences in photosynthetic light-harvesting complexes can be attributed to low effective temperatures in the relevant molecular systems, which can be due to strong coupling with low frequency vibronic modes. This reasoning could also be applicable here; whether such a strongly coupled mode exists remains to be investigated. We  assume that the value of $|\langle S_z\rangle|$ for the Fe$^{2+}$ and Q$^-$ spin system lies anywhere between 0 and 2.5. Combined with reasoning in the previous paragraph, we therefore consider values of the effective magnetic field generated by the fast-thermalising spin system to be between 1 and 100G.\\
\\
The reduction in the triplet yield $\Delta\phi_T$ is defined as the difference between the triplet yield without and with effective magnetic field $B_{\mathrm{eff}}$ generated by the Fe$^{2+}$ and Q$^-$ spin system. While it is well-established that large fields (greater than $\sim$100 G) can induce significant changes in radical pair product yields (greater than $\sim$10$\%$), see ref. \cite{steiner89} for a review, it is less clear what effect a magnetic field of comparable strength to electron-nuclear hyperfine interactions may have on radical pair reactions.\\
\\ 
We solve the master equation, Eq. (19), for three cases with an increasing number of nuclei coupled to each radical, and investigate the triplet yield reduction when both the hyperfine interaction $A$ and the interaction with the effective magnetic field $\omega$ lie in the range $1 \leq A,\omega\leq 100$ G.\\
\begin{figure}
\flushleft
  \includegraphics[scale=0.75]{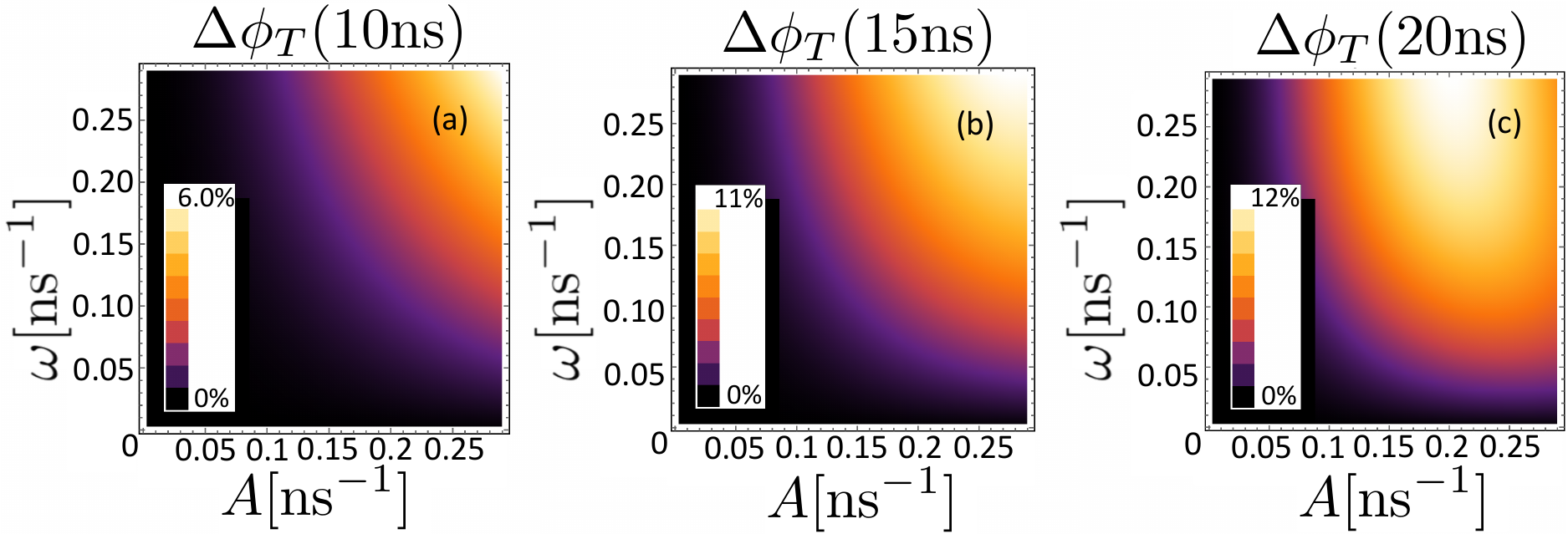}
    \caption{The triplet yield reduction $\Delta\phi_T$ is plotted for $N=1$ as a function of the hyperfine coupling with the single nucleus $A$ and the frequency $\omega$ of the radicals in the effective field $B_{\mathrm{eff}}$, with values of $k_S=k_T=k=1/t_f$ for times $t_f$ of (a) 10, (b) 15 and (c) 20 ns. The interradical coupling $J$ as well as the temperature of the environmental nuclei are set to zero.}
\end{figure}
\begin{figure}
\flushleft
  \includegraphics[scale=0.75]{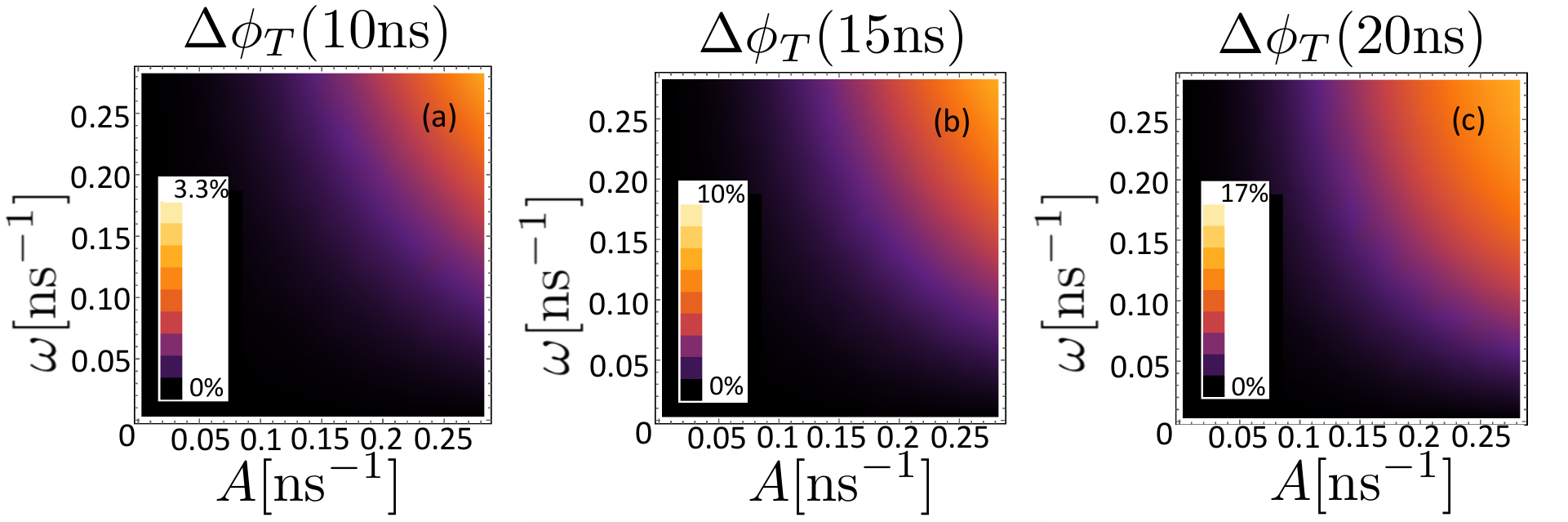}
    \caption{The triplet yield reduction $\Delta\phi_T$ is plotted for $N=3$ as a function of the hyperfine coupling $A$ for a single nucleus coupled to radical 1 and two nuclei coupled to radical 2, and the frequency $\omega$ of the radicals in the effective field $B_{\mathrm{eff}}$. The magnitudes of the hyperfine couplings are given by $\sqrt{N}A_1=\sqrt{N}A_2=A$. The recombinations rates are given by $k_S=k_T=k=1/t_f$ for times $t_f$ of (a) 10, (b) 15 and (c) 20 ns. The interradical coupling $J$ as well as the temperature of the environmental nuclei are set to zero.}
\end{figure}
\begin{figure}
\flushleft
  \includegraphics[scale=0.75]{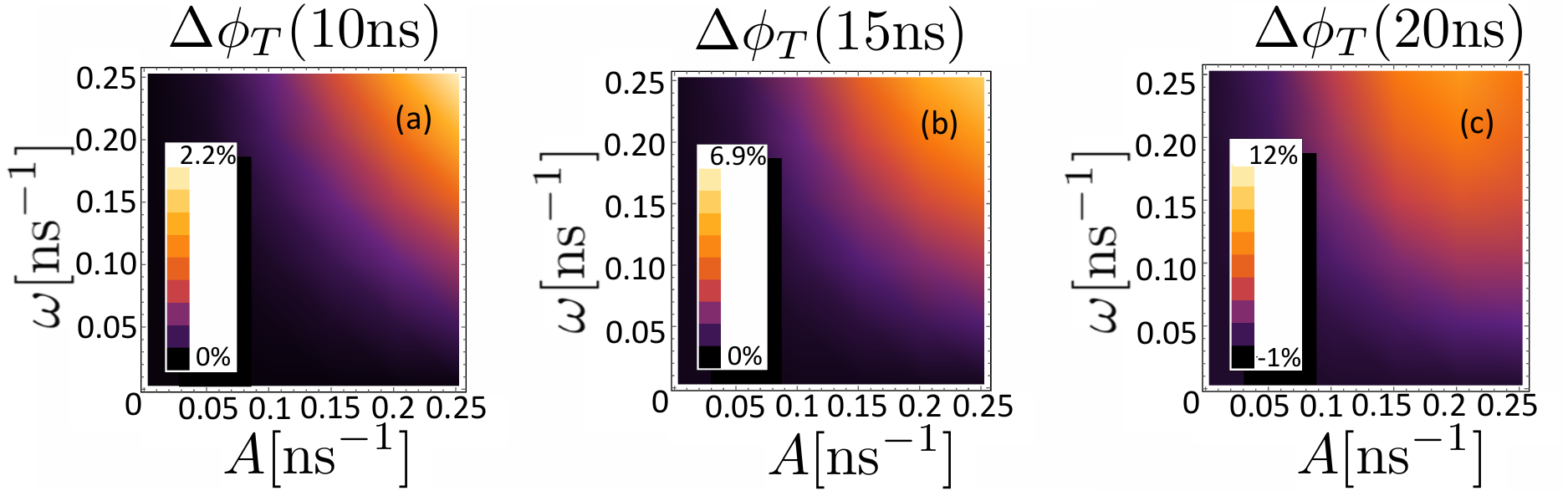}
    \caption{The triplet yield reduction $\Delta\phi_T$ is plotted for $N=5$ as a function of the hyperfine coupling $A$ for three nuclei coupled to radical 1 and two nuclei coupled to radical 2, and the frequency $\omega$ of the radicals in the effective field $B_{\mathrm{eff}}$. The magnitudes of the hyperfine couplings are given by $\sqrt{N}A_1=\sqrt{N}A_2=A$. The recombinations rates are given by $k_S=k_T=k=1/t_f$ for times $t_f$ of (a) 10, (b) 15 and (c) 20 ns. The interradical coupling $J$ as well as the temperature of the environmental nuclei are set to zero.}
\end{figure}
\\
In Figs 3, 4 and 5, the triplet yield reduction $\Delta\phi_T$ is plotted for $N=1,3$ and 5, respectively, as a function of the hyperfine coupling with the nuclei $A$ and the frequency $\omega$ of each radical in the effective field $B_{\mathrm{eff}}$, with values of $k_S=k_T=k=1/t_f$ for times $t_f$ of (a) 10, (b) 15 and (c) 20 ns. We scale the magnitude of the hyperfine couplings as $1/\sqrt{N}$, i. e. $\sqrt{N_1}A_1=\sqrt{N_2}A_2=A$ and set the interradical coupling $J$ as well as the temperature of the environmental nuclei to zero as a first approach to determine the extent of the protective effect.\\
\\
It can be seen that the extent of the protective effect is greatest for the case $N$=3, with a maximum triplet yield reduction of $\Delta\phi_T^{\mathrm{max}}$=17$\%$. The minimum of $\Delta\phi_T^{\mathrm{min}}$=-0.60$\%$ occurs for $N=5$ for $k=0.05$ ns$^{-1}$.\\
\\
The so-called low field effect can lead to a boost in the concentration of free radicals for singlet geminate radical pairs, through an increase in triplet product yield when $\omega$ is comparable to or smaller than $A$ \cite{timmel98}. Timmel et al. note that the anisotropy of hyperfine interactions would reduce degeneracies and therefore the extent of the low field effect. Another factor influencing the extent of the low field effect is the lifetime of the radical pair: Timmel et al. note that low field effects become more unlikely for short (less than $\sim$10 ns) radical pair lifetimes.\\
\\
In Fig. 6, we investigate conditions under which the low field effect occurs for the case $N$=1. It can be seen in Fig. 6 (a) that the triplet yield increases for approximately $0<k\leq 0.5A$, where $A=\omega=50$ G = 0.14 ns$^{-1}$. Setting $k=0.2A=0.028$ ns$^{-1}$ with the same values for $A$ and $\omega$, it can be seen in Fig. 5 (b) that the low field effect occurs when $\omega$ is almost an order of magnitude smaller than $A$, and then \textit{only} for times much longer than the lifetime of the radical pair.\\
\begin{figure}
\centering
  \includegraphics[scale=0.5]{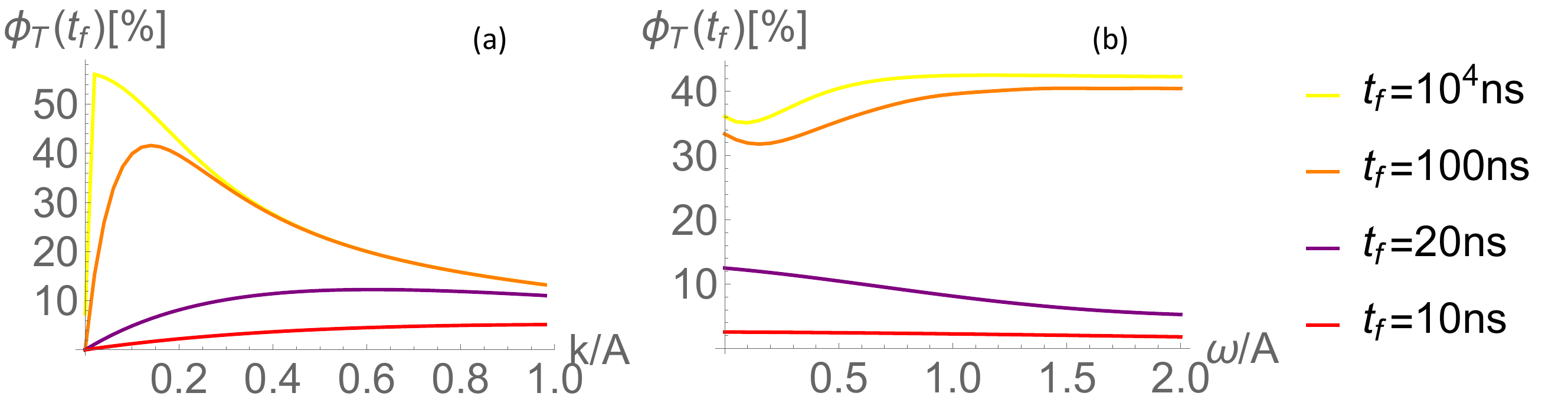}
    \caption{For $k_S=k_T$, it can be seen in (a) that the low field effect occurs for $0<k\leq 0.5A$ for $A=\omega=50$ G = 0.14 ns$^{-1}$. Setting $k=0.2A=0.028$ ns$^{-1}$, it can be seen in (b) that the increase in triplet yield when $\omega$ and $A$ are comparable only occurs for times much longer than the lifetime of the radical pair.}
\end{figure}
\\
To confirm that the low field effect is not significant over the relevant parameter range for the cases N=1, 3 and 5, we consider values of the variables A$_1$, A$_2$ and $\omega$ ranging from 1 to 100 G. We then test the robustness of the effect by varying the recombination rates $k_S$ and $k_T$, the intraradical coupling $J$ and finally the parameter $\alpha\beta$, where $\alpha$ is the energy splitting of the two-level nuclei and $\beta$ the inverse temperature of the nuclei.\\
\\
Setting $k_S=1/15$ ns and defining $k_T=kk_S$, we vary the proportionality constant $k$ between 1 and 10, $J$ between -10 G and 10 G, and the parameter $\alpha\beta$ between $10^{-5}$ and $10^2$. We find that over all considered parameters in the cases $N=1$ and 3, the low field effect is absent  to three significant figures, i.e. the minimum triplet yield reduction $\Delta\phi_T^{\mathrm{min}}$=0.00.\\
\\ 
With an increased number of nuclei in the case $N=5$, there are more degenerate states of the radical pair that are perturbed even by a weak magnetic field which can result in the low field effect. We find that in the case $N$=5 there are sets of parameters in the considered range where an increase in the triplet yield occurs, with a minimum $\Delta\phi_T^{\mathrm{min}}$=-0.60$\%$. However, these results constitute just 7.6$\%$ of the total results.\\
\\
We propose that the high-spin Fe$^{2+}$ ion generates an effective magnetic field, thereby serving the protective function of reducing the triplet product yield in the purple bacterial RC under conditions when forward electron transfer is blocked. Our analysis of the resulting triplet yield reduction has shown that the protective mechanism is significant in its extent, with reductions of up to 17$\%$ for realistic parameters, and, importantly, is robust over the considered parameter ranges. 
\subsection*{Discussion}
Early anoxygenic photosynthetic bacteria probably used Fe$^{2+}$ amongst other reductants to produce organic material from carbon dioxide \cite{pierson94}. Subsequent atmospheric oxygenation \cite{heinrich06} was a precursor to one of the most significant extinction events in Earth's history, and the organisms that survived developed mechanisms of protection against potentially destructive reactive oxygen species. Whether the large spin of the Fe$^{2+}$ ion positioned between the two ubiquinones in both bacterial and photosystem-II RCs plays a functional role has remained unclear, especially since electron transfer dynamics are restored when it is replaced with paramagnetic Zn$^{2+}$ \cite{kirm86,liu91}.\\
\\ 
Based on several experiments \cite{hoff77,blank77,kirm86} that have found the yield of triplet product states from the radical pair P$^{+}$H$^-$ to increase in RCs with the ubiquinone electron acceptor Q$_A$ reduced and the Fe$^{2+}$ ion removed, we show that a fast-relaxing high-spin ion generates an effective magnetic field, thus suppressing triplet state formation as in the so-called radical pair mechanism.\\
\\ 
Electron transfer in photosynthesis proceeds successfully almost 100$\%$ of the time. However, when light-induced damage to the photosynthetic apparatus requires repair, radical pair recombination can occur, and the instantaneous spin multiplicity of the radical pair determines the spin state of the products. Triplet product states can react with molecular oxygen forming highly reactive singlet oxygen. Evidence shows that singlet oxygen inhibits the repair of light-induced damage in photosystem II \cite{hakala11}, with damage occurring for every 10-100 million photons intercepted \cite{eckert91,tyy96}, which corresponds to the entire reaction centre having to be taken apart for repair every half hour in normal sunlight \cite{vang11}. Interestingly, it has also been found that static magnetic fields have significant effects on plant growth \cite{alad03,florez07,vash10}.\\
\\
Given that radical pair recombination does occur under natural conditions, and based on experimental observation of the relative magnetic field effect in bacterial RCs with and without the Fe$^{2+}$ ion, we propose that the large spin of the Fe$^{2+}$ ion plays a protective role by contributing towards preventing an event potentially lethal to the cell: singlet oxygen production. To demonstrate this protective mechanism, we consider a simple model of a system consisting of the radical pair, the thermalised Fe$^{2+}$ ion and coupled reduced ubiquinone molecule Q$_A^-$, in cases with one, three and five spin-1/2 nuclei coupled to the radical pair. Solving the master equation for the system, we find that the protective effect is robust for the range of parameters relevant for the system, with triplet state reductions due to the large spin as high as 17$\%$.\\
\\
There exists a body of literature, see ref. 68 for a review, showing how a weak magnetic field can affect radical recombination reactions in an opposite way to the `normal' radical pair mechanism, namely by reducing rather than enhancing the singlet yield. However, we find that for the parameter range and timescale relevant for the bacterial reaction centre, the relatively weak effective magnetic field generated by the Fe$^{2+}$ ion decreases the triplet yield in 92$\%$ of considered cases, with a maximum increase of 0.60$\%$ in the few remaining cases.\\
\\
The water-splitting photosystem II RC is structurally and functionally homologous to the bacterial RC \cite{michel88}: the cofactors are bound in two-fold symmetry, electron transfer occurs along the active $D_1$ branch and the radical pair P$_{D_1}^{+}$H$_{D_1}^{-}$, formed by a chlorophyll $a$ and pheophytin $a$ molecule, gives rise to triplet states in blocked RCs \cite{rutherford81,groot94}. It has been observed that the quinone-iron complex has an influence on radical pair recombination in photosystem II \cite{mieghem89}. We suggest that the effective magnetic field generated by the Fe$^{2+}$-Q${^-}$ spin system plays an analogous protective role in photosystem II to the bacterial RC, with further investigation into this proposal currently underway.\\
\\
Based on experimental observations that the high-spin Fe$^{2+}$ ion affects photosynthetic radical pair reactions, we propose that spin plays a direct role in contributing towards the prevention of destructive events in photosynthetic RCs. This work motivates further experimental study of the role of the Fe$^{2+}$ ion in both bacterial and photosystem II RCs, as well as raising the interesting question of whether the effective magnetic field generated by a fast-thermalising spin may play a role in other biological processes, particularly those where magnetic field effects have been observed \cite{vanag88,grissom95,galland05}. Furthermore, the robustness of the effect informs a useful design principle for artificial photosynthetic systems also subject to triplet-induced damage, namely, that a large spin in the vicinity of a radical pair can reliably and significantly reduce triplet product formation.

\section*{Acknowledgements}
For A. M., I. S. and F. P., this work is based upon research supported by the South African Research Chair Initiative of the Department of Science and Technology and National Research Foundation. R. v. G. was supported by the Royal Dutch Academy of Sciences (KNAW), the VU University Amsterdam, TOP grant (700.58.305) from the Foundation of Chemical Sciences part of NWO and the advanced investigator grant (267333, PHOTPROT) from the European Research Council.
\section*{Author contributions}
All authors A. M., I. S., F. P. and R. v. G. were actively involved in discussions and formulation of the theory. A. M. and I. S. did the calculations and numerics. A. M. wrote the paper, and all authors A. M., I. S., F. P. and R. v. G. contributed extensively to its finalisation.
\section*{Competing financial interests statement}
The authors have no competing interests as defined by Nature Publishing Group, or other interests that might be perceived to influence the results and/or discussion reported in this paper.

\end{document}